\documentclass[aps,twocolumn]{revtex4}
\usepackage{amssymb}

\newtheorem{theorem}{Theorem}
\newtheorem{acknowledgement}[theorem]{Acknowledgement}

\input{tcilatex}

\begin{document}

\title{Femtosecond ''snapshots ''of gap--forming charge-density-wave
correlations in quasi-two-dimensional dichalcogenides 1$T$-TaS$_{2}$ and 2$H$%
-TaSe$_{2}$.}
\author{J.Demsar$^{1,2},$ L.Forr\'{o}$^{3}$, H. Berger$^{3}$, and D.
Mihailovic$^{1}$}
\affiliation{$^{1}$Jozef Stefan Institute, Jamova 39, SI-1001 Ljubljana, Slovenia\\
$^{2}$Los Alamos National Laboratory, MS K764, Los Alamos, NM 87545, USA\\
$^{3}$Physics Dept., EPFL CH-1015 Lausanne, Switzerland.}

\begin{abstract}
Time-resolved optical spectroscopy of collective and single-particle
excitations of 1$T$-TaS$_{2}$ and 2$H$-TaSe$_{2}$ reveals the presence of a
large gap in the excitation spectrum on the femtosecond timescale,
associated with the formation of various degrees of CDW\ order. In common
with superconducting cuprates, excitations with energies less than the full
gap show much slower relaxation. This separation of timescales cannot be
explained in a quasi-2D Fermi-Liquid picture with an anisotropic gap but
rather suggests the formation of a fluctuating spatially inhomogeneous state
eventually forming a long-range ordered state at low temperatures.
\end{abstract}

\maketitle

Dimensionality can have quite a profound effect on the ground state
properties of materials. For example quasi-one-dimensional metals often
undergo a Peierls-distortion to become insulating at low temperatures\cite%
{Gruner}, or form strange ''Luttinger'' metals in which collective
excitations give rise to peculiar low-temperature properties\cite{Luttinger}%
.\ The ground state of two-dimensional (2D) materials in some cases is also
very peculiar. Quasi-2D charge-density wave (CDW) dichalcogenides have been
receiving renewed attention recently, particularly because they are thought
to exhibit some important similarities to the high-temperature
superconducting cuprates (HTSC). Both are layered, highly anisotropic
materials which are often described in terms of a quasi-2D Fermi surface
(FS) in their normal state. In HTSCs, it is commonly believed that the
superconducting gap has nodes along certain directions on the FS due to the $%
d$-wave component of the order parameter, whereas in 2D-CDW systems a CDW\
gap is also expected only along certain wavevectors, remaining gapless (and
metallic) on other regions of the FS. The low-energy single particle
excitations in the two classes of compounds might therefore be expected to
show some important common features related to reduced dimensionality.
However, the validity of the Fermi-liquid (FL) concept when applied to
low-temperature properties in HTSCs has repeatedly been brought into
question, suggesting that new insight into the physics of quasi-2D systems
may be gained by investigating the low-energy electronic gap structure and
carrier recombination dynamics of quasi-2D CDW dichalcogenides with
femtosecond spectroscopy. The time-resolved technique has been shown to give
femtosecond\ ''snapshots'' of the low-energy gap structure and as such
presents an excellent alternative viewpoint compared to the more usual
time-averaging frequency-domain spectroscopies\cite{Kabanov,Slow,TRbb}.

Here we apply the technique to the study of single-particle (SP) and
collective excitations in two different quasi-2D CDW dichalcogenides 1$T$-TaS%
$_{2}$ and 2$H$-TaSe$_{2}$. We focus on the issue of single particle
dynamics and gap formation in the two materials (note that the effective
shutter speed is on the femtosecond timescale) and compare the results with
the predictions based on a quasi-2D FL\ picture, finding some fundamental
discrepancies between the expected behaviour and our observations.

Both 1$T$-TaS$_{2}$ and 2$H$-TaSe$_{2}$ exhibit\ a series of successive
phase transitions, starting from a highly anisotropic quasi-2D metallic
state at high temperatures, and ending with a commensurate (\textit{c-}) CDW
state at low temperatures \cite{Review}. (A summary of the sequence of
transitions is shown in the inserts to Fig.1.) At room temperature 1$T$-TaS$%
_{2}$ is in a nearly-commensurate (\textit{nc}-) CDW phase (Fig.1a). Around $%
T_{nc-c}=200$ K it undergoes a strongly first-order ''lock-in'' transition
to a \textit{c}-CDW state \cite{Review}. In spite of the expected appearance
of a gap in parts of the Fermi surface due to nesting at $T_{nc-c}$, the
whole FS was found to exhibit a ''pseudogap'' feature already at room
temperature with a finite density of states (DOS) at $E_{F}$ \cite%
{DardelPilloManske}. Upon lowering the temperature, a further \emph{abrupt}
decrease in the DOS is observed near $E_{F}$ at $T_{nc-c},$ accompanied by
an order of magnitude \textit{increase }in resistivity. Yet in spite of the
presence of a CDW\ gap at low temperatures 1$T$-TaS$_{2}$ is reported to
have a small but finite DOS at $E_{F}$ in the low-temperature \textit{c}%
-phase \cite{DardelPilloManske}.

2$H$-TaSe$_{2}$ on the other hand is expected to bear close resemblance to
cuprates. It exhibits metallic properties above room temperature (Fig.1b).
Upon cooling it undergoes a second order phase transition to an
incommensurate (\textit{i}-) CDW state at $T_{n-i}=122$ K. This phase
transition is reportedly accompanied by the appearance of a gap on the Fermi
surface (FS) centered at the K point, but apparently - according to
photoemission studies\cite{Liu} - remains gapless on the part of the FS
centered at the $\Gamma $ point. The transition is accompanied by a decrease
in the scattering rate and a corresponding \emph{drop} in resistivity \cite%
{Vescoli}. The onset of a \textit{c}-CDW phase at $T_{i-c}=88K$ leaves the
excitation spectrum as well as the transport and thermodynamic properties
almost unaffected \cite{Review,Liu}. Upon warming from the $c$-phase, an
additional ''striped'' incommensurate (\textit{si}-) phase has been reported
between 92 and 112 K \cite{Fleming}. \ 

The experiments reported here were performed on freshly cleaved single
crystals, using a pump-probe set-up with a mode-locked Ti:Sapphire laser (50
fs pulses at 800 nm) for\ both pump and probe \thinspace pulse trains. The
photoinduced change in reflectivity $\QTR{sl}{\Delta }\mathcal{R}/\mathcal{R}
$ was measured using a photodiode and lock-in detection. The pump laser
power was kept below 5 mW and the pump/probe intensity ratio was $\sim $100.
The steady-state heating effects were accounted for as described in \cite%
{ACS98} giving an uncertainty in temperature of $<2$ K. As the optical
penetration depth is $\gtrsim $ 100 nm, the technique is essentially a 
\textit{bulk} probe, and since we are using very weak photoexcitation \cite%
{Kabanov} the system as a whole remains close to equilibrium.

The sequence of relaxation events after photoexcitation is common to metals
(incl. superconductors) and CDW materials \cite{Kabanov,Allen,Ippen}: the
photoexcited (PE) carriers first rapidly thermalize by \emph{e-e} scattering
(within $\tau _{e-e}\approx 10$ fs) and then transfer their energy to the
lattice with a characteristic electron-phonon relaxation time given by $\tau
_{e-ph}=k_{B}T_{e}/\hbar (\lambda \langle \omega ^{2}\rangle )$ \cite%
{Allen,Ippen}, where $\lambda $ is the electron-phonon coupling constant,
and $\langle \omega ^{2}\rangle $ is the mean square phonon energy. Using $%
\omega \approx 200$ cm$^{-1}$ \cite{SugaiRaman} and $\lambda \approx 0.3$ %
\cite{Kamarchuk} we obtain (for the excitation intensities used here) $\tau
_{e-ph}\approx 100$ fs which is the time required for the PE carriers to
relax to energies close to $E_{F}$. If a gap in the DOS\ is present near $%
E_{F}$, the resulting relaxation bottleneck causes carriers to accumulate in
states above the gap. The population of these carriers $n(T,t)$ can then be
probed by a delayed probe laser pulse and its time evolution directly gives
the \emph{energy} relaxation time $\tau _{s}$ for their recombination. From
the $T$-dependence of $n(T,t)$ direct information on the gap magnitude, its $%
T$-dependence, as well as something about its anisotropy can be extracted %
\cite{Kabanov}. In addition to this SP response, the perturbation of the
charge density caused by the PE carriers can excite CDW collective modes\cite%
{TRbb,Zeiger} - particularly the amplitude mode (AM) - which can be observed
as an oscillatory response superimposed on the SP relaxation data.

In Fig.1 we show the induced reflectivity $\Delta \mathcal{R}/\mathcal{R}$
in 1$T$-TaS$_{2}$ and 2$H$-TaSe$_{2}$ as a function of time after
photoexcitation at different temperatures. The general feature of the data
in both cases is the appearance of a decaying transient and a superimposed
oscillatory response due to the SP and collective mode relaxations
respectively. The two contributions can be easily deconvolved \cite{TRbb} by
fitting the transient reflectivity signal to a function of the form:

\begin{eqnarray}
\Delta \mathcal{R(}t,T\mathcal{)}/\mathcal{R} &=&A(T)e^{-t/\tau _{A}}\cos
(\omega _{A}t+\phi _{0})+  \nonumber \\
&&S(T)e^{-(t/\tau _{s})^{s}}+D(T).
\end{eqnarray}
\newline
The first term describes the modulation of the reflectivity\ due to coherent
oscillations of the AM mode of frequency $\omega _{A}$, where $\Gamma
_{A}=\pi /\tau _{A}$ is the AM damping constant \cite{Zeiger}. The second
term describes the SP response and the third term describes the
''background'' which is the contribution from decay components with
lifetimes longer than the inter-pulse separation of 10 ns \cite{Slow}. The
signal appeared to be independent of probe polarization in both materials.

Let us first analyze the oscillatory AM response. In 1$T$-TaS$_{2}$ (Fig.
1a)) below $T_{c-nc}$, the frequency of oscillation corresponds closely to
the AM frequency $\omega _{A}$ as determined by Raman spectroscopy. The $T$%
-dependence of $\omega _{A}$ and $\Gamma _{A}$ determined from the fit are
plotted in Fig. 2a) showing excellent agreement with the established Raman
data \cite{SugaiRaman}. $\Gamma _{A}$ strongly increases near $T_{c-nc}$. A
profound hysteresis is observed in the $T$-dependence, where a rather sharp
drop in $\omega _{A}$ (and increase in $\Gamma _{A}$) coincides with $%
T_{c-nc}$ - see inset to Fig. 1a). In contrast, for 2$H$-TaSe$_{2}$ the $T$%
-dependence of $\omega _{A}$ and $\Gamma _{A}$ (Fig. 2b))\ is much more
mean-field like, becoming overdamped $(\omega _{A}\simeq \Gamma _{A})$
around $T=110$ K. (For comparison, Raman data for $\omega _{A}$ and $\Gamma
_{A}$ is also shown \cite{SugaiRaman}.)

Let us now turn to the SP excitations. While in 2$H$-TaSe$_{2}$ the SP
transient can be reproduced well by a \emph{single }exponential decay over
the whole temperature range (i.e. $s=1$), the SP relaxation dynamics in 1$T$%
-TaS$_{2}$ requires a \emph{stretch }exponential decay fit with $s\sim 0.5$
to fit the data adequately. (Although the SP transient in 1$T$-TaSe$_{2}$
can also be fit by the sum of two exponentials, the two components have the
same $T$-dependence, which suggests that we are dealing with a single
relaxation process with non-exponential dynamics rather than two distinctly
independent processes). The observation of a stretch exponential decay -
which typically describes systems with a spread of relaxation times - is
consistent with the observed finite DOS at E$_{F}$ below $T_{c-nc}$ \cite%
{DardelPilloManske}. (Since $\tau \sim 1/\Delta $ [Ref. 3] the observed
stretch exponential decay actually implies a near-Gaussian spread of $%
1/\Delta $.)

The $T$-dependences of the amplitude $S(T)$ and $\tau _{s}$ (using $s=0.5$)
for 1$T$-TaS$_{2}$ and 2$H$-TaSe$_{2}$ are plotted in Figs. 3a) and b)
respectively. In 1$T$-TaS$_{2}$ the relaxation dynamics are clearly strongly
affected by the lock-in transition around 200K. \ We observe an abrupt
hysteretic change of $S(T)$ and $\tau _{s}$ around $T_{nc-c},$ consistent
with an abrupt appearance of a gap at $T_{nc-c}$ suggested by other
experiments \cite{DardelPilloManske}.\ Upon further cooling $S$ remains more
or less constant, while $\tau _{s}$ slowly decreases. Upon warming, a rapid
drop in $S$ and $\tau _{s}$ associated with gap closure now occurs at around
220 K, consistent with the hysteresis observed in the collective mode
response in Fig. 2a). Above~230 K the photoinduced transient is fast and
very weak.

The single-exponential fit to the SP relaxation in 2$H$-TaSe$_{2}$ over the
entire $T$-range is surprising, since below $T_{n-i}$, the system is
expected to have a highly anisotropic gap with gapless regions over parts
the FS\cite{Liu} implying that the relaxation should deviate strongly from a
single exponential. (A similar problem arises in cuprates, where the decay
is also exponential, in spite of the fact that nodes in the gap would
expected to give rise to strongly non-exponential decay\cite{Kabanov}.) The $%
T$-dependence of $S(T)$ and relaxation time $\tau _{s}$ obtained from the
single exponential fit to the data on 2$H$-TaSe$_{2}$ are plotted in Fig. 3.
Below $\sim $ 120 K both $S(T)$ and $\tau _{s}(T)$ can be very well
described by model for carrier relaxation across a well-defined
temperature-dependent gap. The theoretical fit for $S(T)$ and $\tau
_{s}\propto 1/\Delta (T)$ using the model by Kabanov et al. \cite{Kabanov}
using a BCS gap function is superimposed. The value of the gap from the fit $%
2\Delta (0)=70\pm 10$ meV is somewhat smaller than the maximum gap obtained
from tunneling \cite{tunelling}, but in good agreement with the maximum gap
value of 65 meV from ARPES\cite{Liu}. The divergence of $\tau _{s}$ and the
drop in $S(T)$ as $T\rightarrow T_{i-n}\approx 120$ K from below are
unmistakable signs of a $T$-dependent gap which is closing near $T_{i-n}$ %
\cite{Kabanov}. Above $\sim 140$ K the relaxation time and $S(T)$\ show only
a weak $T$-dependence with $\tau _{s}\sim 0.1-0.17$ ps, close to the
estimated $\tau _{e-ph}$. $\tau _{e-ph}$ is expected to increase with
increasing temperature (see Fig. 3 d) and laser power \cite{Allen}, which is
indeed observed: at room temperature a $\sim $30 \% increase in $\tau _{s}$
is observed at a 3-fold increase in the pump fluence - apparently confirming
that the relaxation above $T\sim 140$ K is primarily due to $e-ph$
thermalization \cite{Ippen}. On the other hand, the behavior of $S(T)$\ just
above $T_{i-n}$ is rather unusual, showing a rapid increase with increasing
temperature between $120$ and $\sim 140$ K. This could be attributed to the
presence of segments of ordered CDW \cite{TRbb} giving rise to a negative $%
\Delta R/R$ transient\cite{YBCO124} with vanishing amplitude at $T\gg
T_{i-n} $ in addition to positive transient due to $e-ph$ thermalization.
Since the two relaxation times are comparable, adding up the two would
result in the behavior just as observed in Fig. 3b. It should be noted that
similar behavior was found also in K$_{0.3}$MoO$_{3}$ \cite{TRbb}.

A very important feature of the data is the slow relaxation component $D(T)$
shown in Fig. 3e) for 2$H$-TaSe$_{2}$. This relaxation is typically
attributed to intragap localized states near $E_{F}$ and its T-dependence
gives independent data on the the $T$-dependence of the gap \cite{Slow}. The
magnitude of $D$\ is very small and more or less $T$-independent in 1$T$-TaS$%
_{2}$, but is very pronounced in 2$H$-TaSe$_{2}$ with a $T$-dependence
typical for a $T$-dependent gap\cite{Slow}. Comparing with cuprates, the
behaviour of 2$H$-TaSe$_{2}$ is similar to overdoped YBa$_{2}$Cu$_{3}$O$%
_{7-\delta }$\cite{Over}. The model in ref. \cite{Slow} predicts $D(T)$ to
be proportional to $\Delta (T)^{-3/2}$ at low temperatures (regardless of
the gap anisotropy) which is in good agreement with the observed $T$%
-dependence of the fast SP lifetime $\tau _{s}$, amplitude $\ S(T)$ as well
as $\nu (T)$. A fit to the data using $\Delta (T)$ of BCS form is shown as a
solid line in Fig. 3e) for 2$H$-TaSe$_{2}$.

The emerging picture based on the time-domain measurements on 2H-TaSe$_{2}$
presented here is one in which the low-temperature state shows \emph{a clear
large gap }in the excitation spectrum on the femtosecond timescales (not
just a depression in the DOS such as is observed in time-averaged
experiments). There is also clear evidence for very slow relaxation of
excitations whose energy is less than the maximum gap. The observed
behaviour is in clear contradiction with a FL interpretation, where the SP\
relaxation would be expected to occur primarily in the gapless regions of
the FS (in the nodes for the case of superconductors). The observation of
only a large SP gap on the femtosecond timescale implies that there are
certain momenta associated with the gapless regions which are either
inaccessible to quasiparticles, or - implying a breakdown of the FL picture
alltogether - simply that \emph{extended} states with these wavevectors do
not exist\emph{\ }at all. The latter behaviour is consistent with the notion
of fluctuating \emph{locally} \emph{ordered }regions in space, in which case
it becomes clear why one cannot speak of FL-like quasiparticle excitations
with well defined momenta. The precursor ''pseudogap state'' appears to be
associated with the fluctuating presence of fully gapped short-range-ordered
CDW patches or segments, similar to the locally gapped regions in real space
arising from a statistically fluctuating population of pre-formed pairs in
HTSCs\cite{Kabanov,Over}. We can also remark on the striking similarity in
the SP relaxation dynamics - and gap structure - between 2$H$-TaSe$_{2}$
below $T_{n-i},$ and overdoped and optimally doped HTSC\cite{Over} below $%
T_{c}$. In both cases the SP relaxation dynamics is governed by the opening
of a mean-field-like gap, characteristic of a $2^{nd}$ order phase
transition below $T_{n-i}$ (or $T_{c}$). The observed divergence of the
relaxation time and $T$--dependence of $S(T)$\ are unmistakeable signatures
of such a gap indicating that \emph{a collective }mechanism is responsible
for joining the short-range local correlated segments into a long-range
ordered CDW (or superconducting)\ state. This behaviour is quite different
to the more glass-like 1$T$-TaS$_{2}$ with its 1$^{st}$ order behaviour and
stretch exponential relaxation.

These time-resolved experiments show that irrespective of the fundamental
underlying cause for the instability, these quasi-2D materials, in common
with HTSCs show a transition from a high-temperature uniform metallic state
to a low-temperature correlated state via the formation of a dynamically
inhomogeneous intermediate state in which local precursor CDW segments (or
pairs in the case of HTSCs) appear on the femtosecond timescale. The
time-averaged response (such as is observed in ARPES or infrared spectra)
may then be thought of as the superposition of the different components in
the inhomogeneous state, while the observed anisotropy reveals the
directionallity of the interaction which leads to the formation of long
range order.

Figure 1. The transient reflection $\Delta R/R$\ from a) 1$T$-TaS$_{2}$\ and
b) 2$H$-TaSe$_{2}$\ at a number of temperatures above and below T$_{{nc-c}}$%
\ and $T_{{n-i}}$\ respectively. The signals are offset for clarity. Insets:
The phase diagrams of\ bulk 1$T$-TaS$_{2}$\ and 2$H$-TaSe$_{2}$.

Figure 2. The amplitude mode frequency $\nu _{A}$ (full circles) and damping
constant $\Gamma _{A}=1/(\pi \tau _{A})$ (open circles) as a function of
temperature for a) 1T-TaS$_{2}$ and b) 2$H$-TaSe$_{2}$. The squares
represent the Raman data \cite{SugaiRaman}. $\nu _{A}$ and $\Gamma _{A}$
show hysteresis near $T_{c-nc}$ in 1$T$-TaS$_{2},$ shown on an expanded $T$%
-scale in the inset to panel a).

Figure 3. The $T$-dependence of fast component amplitude, $S(T)$, for a)
1T-TaS$_{2}$ and b) 2$H$-TaSe$_{2}$. In panels c) and d) the $T$-dependences
of the corresponding relaxation times (a stretch exponential was used in
case of 1$T$-TaS$_{2}$). Open symbols: data taken upon warming, solid
symbols: data taken upon cooling. e) The $T$-dependence of the slow
component amplitude. Solid lines in b), d) and e) are fits to the data - see
text.

\begin{acknowledgement}
We would like to acknowledge valuable discussions with V.V. Kabanov.
\end{acknowledgement}

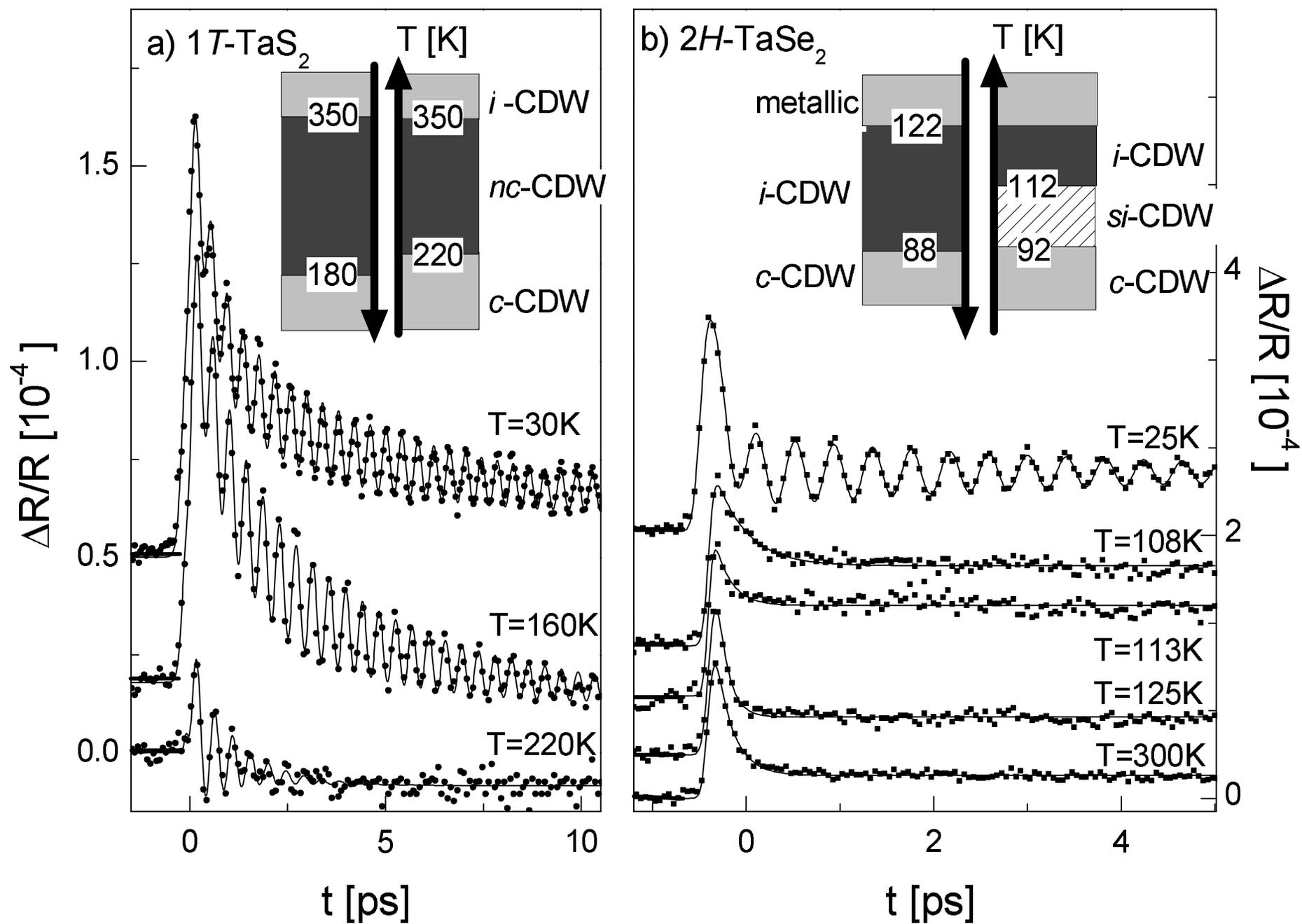
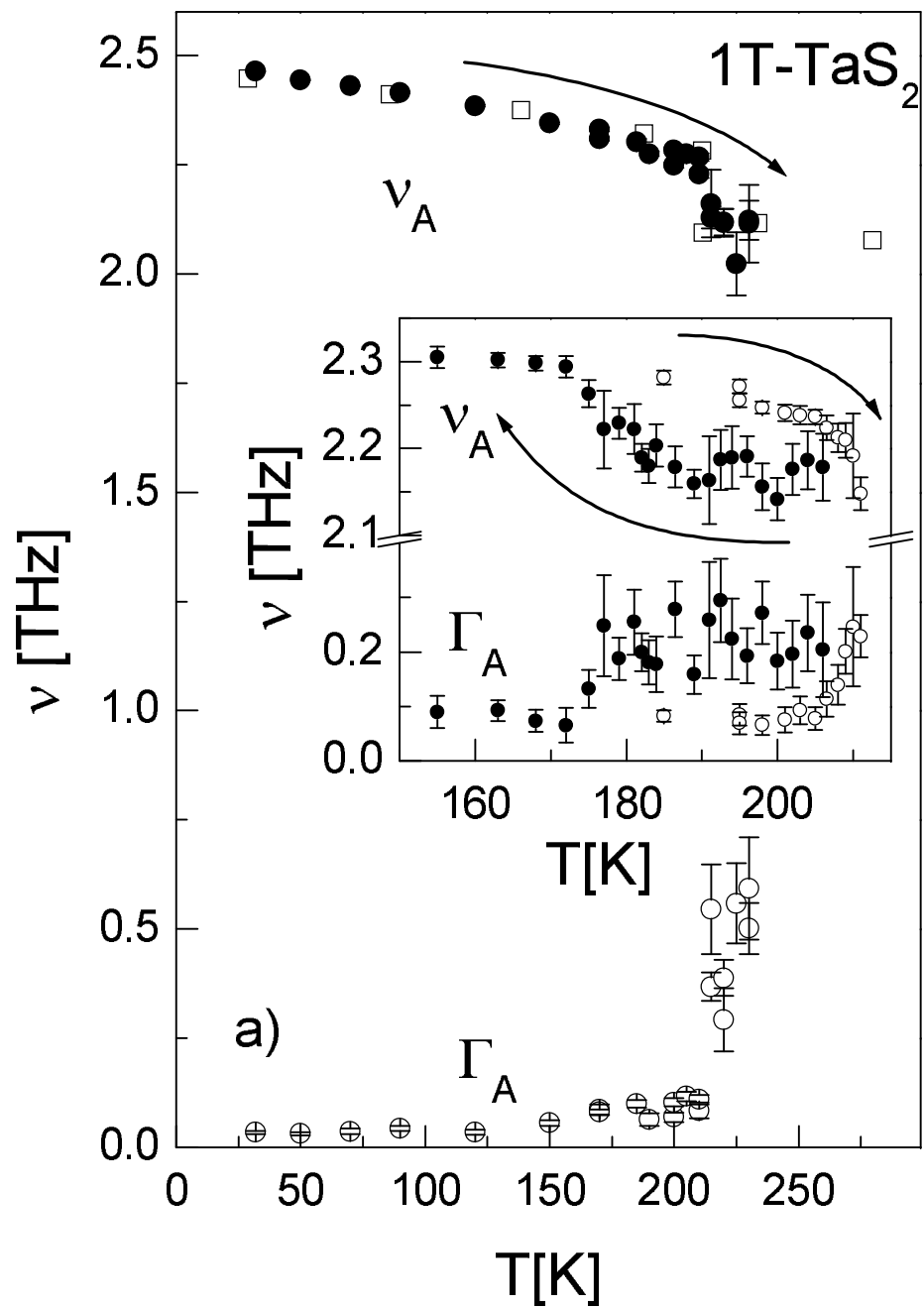
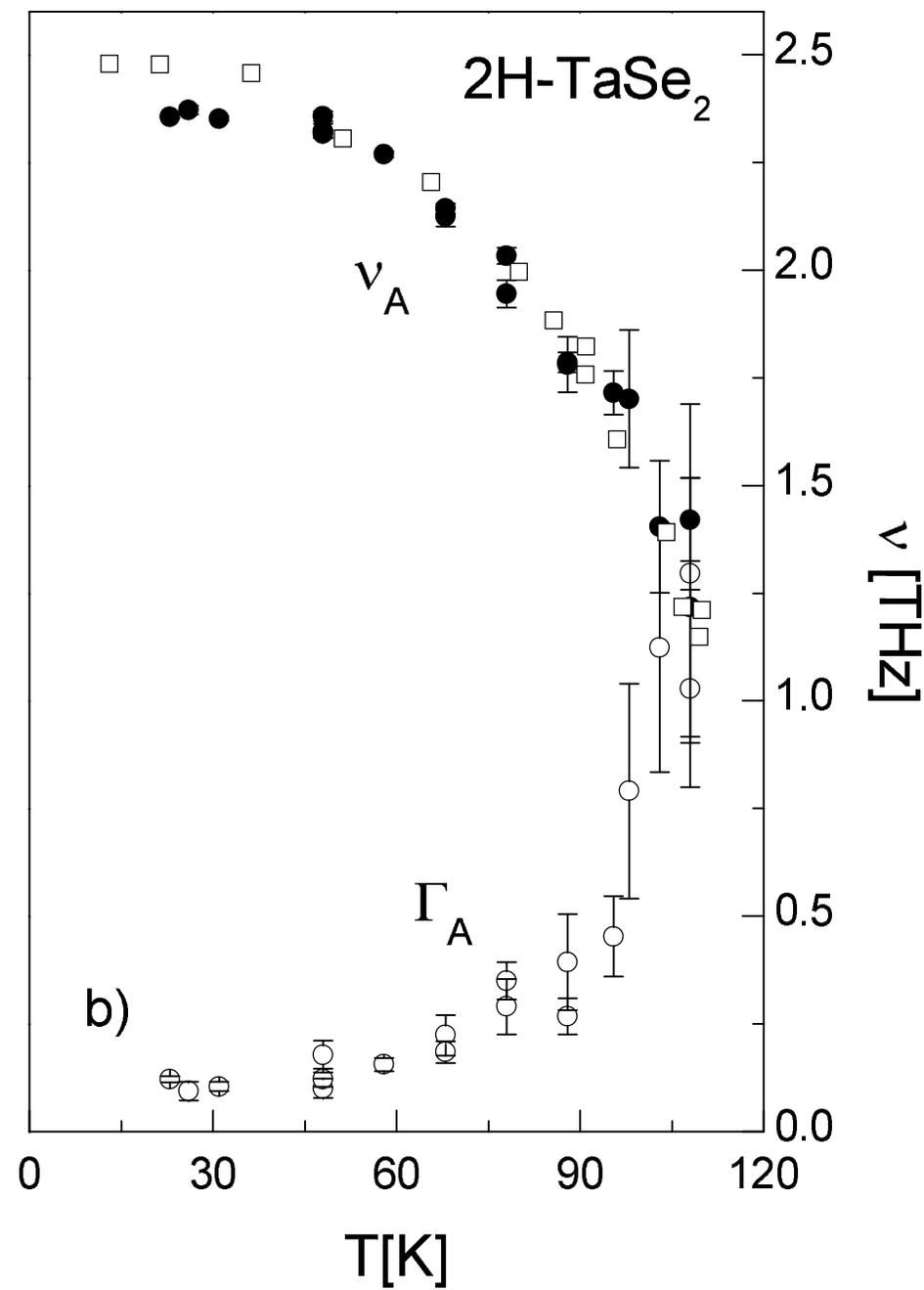
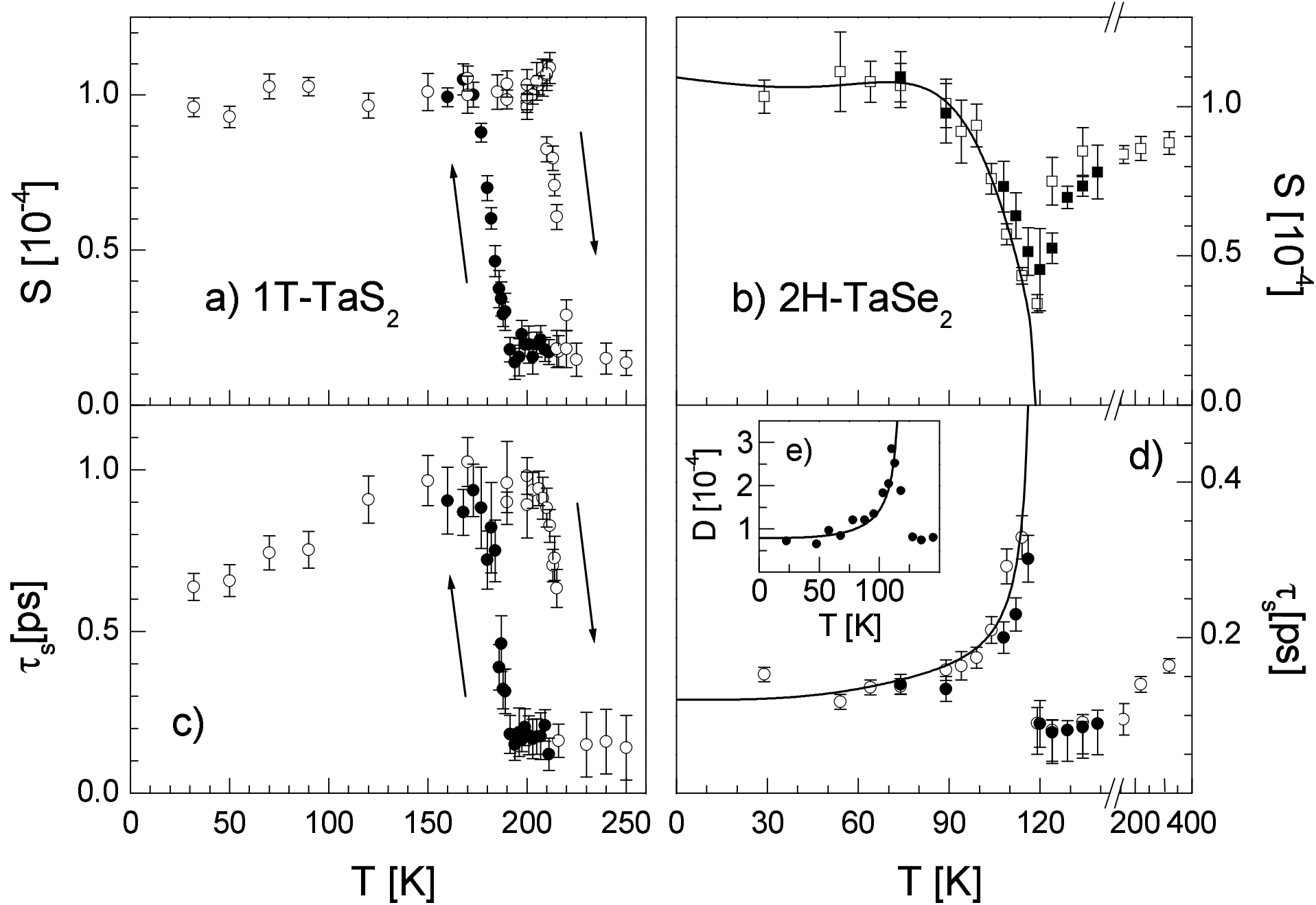
\end{document}